\documentclass{icrc29}
\usepackage{epsfig}
\usepackage{units}
\usepackage{amsmath,amssymb,amstext}
\usepackage{graphicx}
\usepackage{wrapfig}
\begin{document}
%
\title[Measurement of the LDF of UHECR 
Air Showers with the Pierre Auger Observatory]
{Measurement of the Lateral Distribution Function of UHECR 
Air Showers with the Pierre Auger Observatory}
\author[D. Barnhill et al. for the Pierre Auger Collaboration] 
{D. Barnhill, P. Bauleo, M.T. Dova, J. Harton, R. Knapik, J. Knapp, J. Lee, 
\newauthor
M. Mance\~{n}ido, A.G. Mariazzi, I.C. Mari\c{s}, D. Newton, M. Roth, T. Schmidt, 
\newauthor  
A.A. Watson for the Pierre Auger Collaboration}

\presenter{Presenter: P.~Bauleo (bauleo@lamar.colostate.edu),  usa-bauleo-PM-abs2-he14-poster}

\maketitle

%
\vspace*{-2mm}
\section{Introduction}
\label{sec-intro}
The Pierre Auger Observatory~\cite{nimpaper} 
is being used to study cosmic rays with energies larger than
$\unit[10^{19}]{eV}$ with unprecedented
precision and  statistics.  An essential quantity that must 
be deduced from data is the lateral distribution function (LDF) that describes
the decreasing of the signals in the water-tanks as a function of distance.
Knowledge of the LDF is important for the reconstruction of the shower
core and the shower direction. It can also be compared with
model calculations to give useful information relating to primary mass.  
Here we describe how the LDF is measured using the large sample of events
recorded with the surface detector (SD) array and with a small sample
observed with the fluorescence detectors (FD).  For hybrid events, in which SD and
FD measurements of the same shower are available, 
the core position is much better constrained than for SD-only events,
thus providing an important cross-check on the LDF determined from SD measurements alone. 
\vspace*{-2mm}
\section{The Fitting Method}
\label{sec-fit}
The water-Cheren\-kov detectors provide information about Cheren\-kov
photons, which are produced when charged particles cross the tanks. 
The number of Cherenkov photons collected is to a good approximation proportional
to the energy deposit in the tank. 
The signal is calibrated in units of {\em vertical equivalent muons} (VEM) \cite{VEM}. 
The energy deposit, however, depends strongly on the particle type
and the conversion from the Cherenkov signal back to the number of
particles in the tank is not obvious.
For large tank signals ($>\unit[15]{VEM}$) this is not crucial since the
uncertainty $\sigma(S)$ 
of a signal $S$ (in VEM) was determined from data of two detectors positioned 
\unit[11]{meters} apart~\cite{ref:icrc05-ghia} to be $\sigma(S) = 1.06\ \sqrt{S}$.
But for small tank signals
the number of effective particles, $n$, is needed since their
Poissonian fluctuation dominates the uncertainty of the signal
and is required for the maximum likelihood fit.
We have introduced a function that gives
$n$ for a measured signal $S(r)$:  $n = P(r, \theta, E, A) \times S(r)$
where the conversion factor P 
%
%
is called the {\em Poisson factor} and is presently assumed to be
independent of the primary energy, $E$, and mass, $A$, for any
distance, $r$, and zenith angle, $\theta$. The factor reflects the different
energy deposits of different secondaries and is determined by simulations.
Finally we set up a maximum likelihood fit
to determine the parameters of a trial LDF functional form
and, at the same time, the position of the shower core,
by comparing each tank signal, with its fluctuations, to the
value expected from the trial function, $S_{th}$.
%
$
L = \prod_i f_\text{P}(n_i,\mu_i) \times
    \prod_k f_\text{G}(n_k,\mu_k) \times
    \prod_l F_\text{sat}(n_l,\mu_l) \times
    \prod_m F_\text{zero}(n_m,\mu_m).
$
%
The individual factors of the likelihood function are determined using the
information of tanks at their respective distance $r$. The Poissonian probability
density, $f_\text{P}(n_i,\mu_i)$, is calculated for small signals
($S_i<\unit[15]{VEM}$). For larger signals the Gaussian approximation is used,
$f_\text{G}(n_k,\mu_k)$. The effective particle number $n$ of a saturated tank 
represents a lower limit of the actual signal 
and we have to integrate $f_\text{G}$ over all possible values larger
than $n$, to estimate the detecting probability of a signal
larger than $n$. In case of tanks without a
signal we have to sum over all
Poissonian probabilities with a predicted particle number $\mu_i$ and actual
effective particle number $n_i\leq 3$. 

\vspace*{-2mm}
\section{LDF Measurements}
\label{ldf-meas}
\begin{figure}
\begin{minipage}[t]{0.47\textwidth}
\centering
\includegraphics[width=\textwidth]{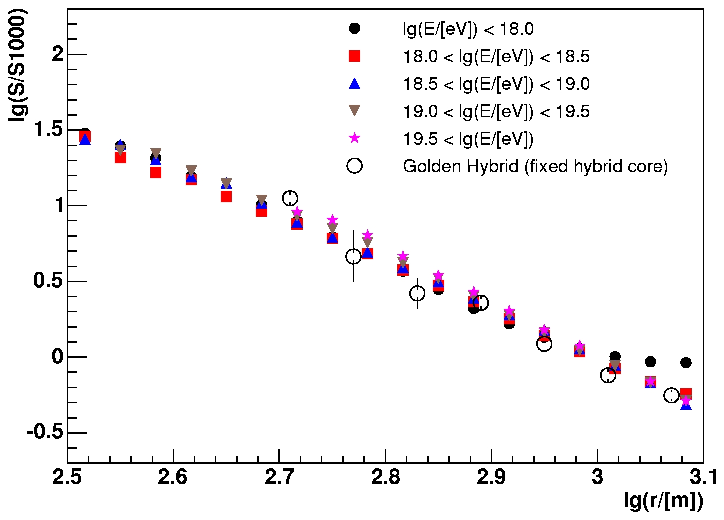}
\caption{\footnotesize 
Averaged LDF for $\sec \theta \in [1.2,1.4]$  (NKG fit and floating
slope).} 
\label{fig:sectheta-plot}
\end{minipage}
\hfill
\begin{minipage}[t]{0.48\textwidth}
\centering
\includegraphics[width=\textwidth]{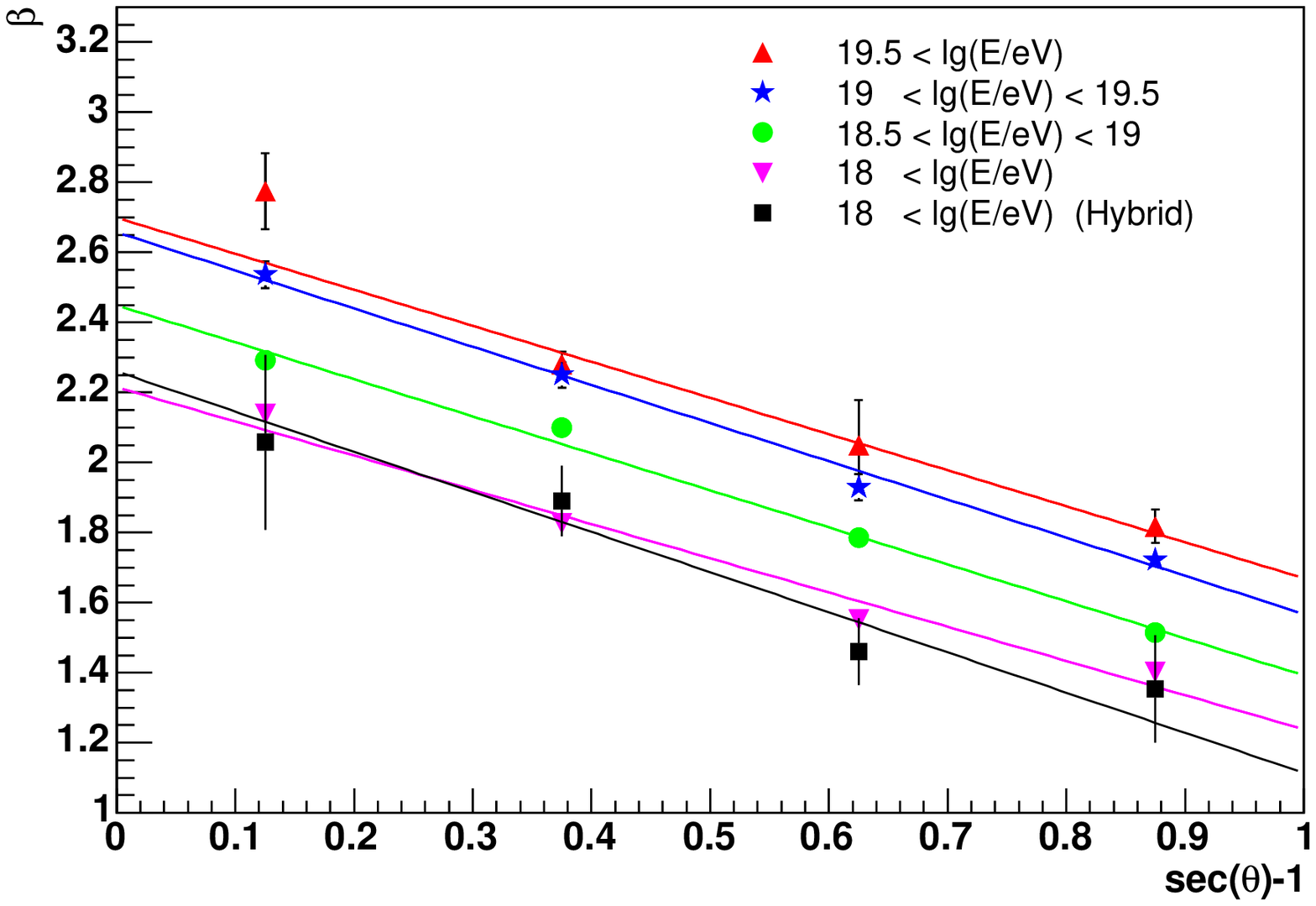}
\caption{Slope $\beta$ vs zenith angle
   as function of energy for SD-only or hybrid geometry. Fits to the slope parameter for different
energy bins are shown} 
\label{fig:beta-vs-secTheta}
\end{minipage}
\end{figure}
In contrast to S(1000) the shape of the lateral distribution does not change
much with energy~\cite{auger-roth}. Therefore, 
the normalisation constant is decoupled
from the shape parameter and showers of different energies are combined. 
The LDF was deduced from experimental data using SD-only and Hybrid events.
%
Data from January 2004 to April 2005 have been used
for the following analysis.  High-quality events have been selected, which had a
successful directional reconstruction with $\theta < 60^\circ$, at least 6
stations with signal above detection
threshold, and a core position well inside the SD array. Using SD-only events 
the following LDFs have been investigated:\quad
(i) a power law: 
$
S(r) = S(1000) \cdot (r/{\rm 1000~m})^{-\nu}
$,
with a $\theta$ dependent index  $\nu = a + b (\sec\theta -1)$, \quad
(ii) a NKG-like function~\cite{ref:nkg}:
$S(r) = A\cdot[(r/r_s)\cdot(1+r/r_s)]^{-\beta}$
with $A = S(1000) \cdot 3.47^{\beta}$,
$\beta = a + b (\sec\theta-1)$ and $r_s =\unit[700]{m}$ (since
  $\beta$ and $r_s$ are strongly correlated, we have fixed $r_s =\unit[700]{m}$ and
  left $\beta$ to vary),
 and \quad
(iii)  a function used by the Haverah Park experiment~\cite{coy}:
$S(r)= k\,r^{-(\eta+r/r_s)} $, if  $r<\unit[800]{m}$, else 
$(\frac{1}{800})^\delta \, k\,r^{-(\eta+r/r_s)+\delta}$ 
with fixed $\delta$, the shape parameter $\eta$ varying with zenith
angle, and $r_s = \unit[4000]{m}$.
These forms were fitted to individual events using a maximum likelihood fit of
core location and LDF at the same time (see section~\ref{sec-fit}). 
\begin{table}[b!]
  \centering
  \scriptsize
  \begin{tabular}{|c|r|r||c|c|c|c|c|c|c|c|c|c|c|c|c|c|}
    \hline
    $\sec\theta$  & \multicolumn{2}{|c||}{number of} &
    \multicolumn{6}{|c|}{NKG } &
    \multicolumn{4}{|c|}{power law} &
    \multicolumn{4}{|c|}{Haverah Park} \\
    range & \multicolumn{2}{|c||}{events} &
    \multicolumn{2}{|c|}{$\beta$ free (hy)} &
    \multicolumn{2}{|c|}{$\beta$ free (sd)} &
    \multicolumn{2}{|c|}{$\beta$ fixed (sd)} &  
    \multicolumn{2}{|c|}{$\nu$ free} &  
    \multicolumn{2}{|c|}{$\nu$ fixed} &  
    \multicolumn{2}{|c|}{$\eta$ free} &  
    \multicolumn{2}{|c|}{$\eta$ fixed}  \\  
    & hy & sd &  m & $\sigma$& m & $\sigma$& m & $\sigma$& m &  $\sigma$& m & $\sigma$& m &
    $\sigma$ & m &  $\sigma$\\
    \hline
    \hline
    $[1.0, 1.2]$ &  5 & 367 &  0.27 & 2.12 & 0.04 & 0.48 &-0.07 & 1.45 & -0.03 & 0.55 & -0.07 & 1.45 & -0.17 & 1.27 &-0.21 & 1.0\\
    $[1.2, 1.4]$ & 14 & 549 & -0.18 & 1.71 & 0.06 & 0.53 &-0.04 & 1.30 & -0.27 & 0.81 & -0.04 & 1.30 & -0.03 & 0.95 & 0.14 & 1.2\\
    $[1.4, 1.6]$ & 17 & 624 & -0.07 & 2.00 & 0.07 & 0.55 & 0.04 & 1.02 & -0.12 & 0.65 &  0.05 & 1.02 & -0.09 & 1.04 & 0.03 & 1.6\\
    $[1.6, 1.8]$ &  8 & 576 & -0.04 & 1.40 & 0.09 & 0.59 &-0.07 & 0.80 & -0.07 & 0.81 & -0.14 & 0.92 &  0.01 & 1.27 & 0.23 & 1.4\\
    $[1.8, 2.0]$ &  6 & 493 & -0.26 & 1.34 & 0.11 & 0.61 &-0.11 & 0.98 & -0.11 & 0.98 & -0.21 & 1.15 & -0.21 & 1.28 &-0.24 & 1.9\\
    \hline
  \end{tabular}
  \caption{\footnotesize Moments (mean, $\sigma$) of residual distribution of exp. data 
    with various LDFs. Only events with $\ge 6$ stations were used in the present analysis.
    For the NKG-like LDF both SD and Hybrid moments are shown, denoted by ``sd'' and ``hy'' respectively.
  }
  \label{tab-res}
\vspace*{-4mm}
\end{table}
Two SD-only analyses were performed.  First, in a four-parameter fit, 
besides the core location x and y, 
the slope parameters $\nu$, $\beta$ and  $\eta$ , respectively,  have been
varied together with the scale factor $S(1000)$. Then a parameterisation of
$\nu$, $\beta$  and $\eta$ as function of $\theta$ was determined, which was then used
in a second analysis fitting only x, y, and $S(1000)$. 
Figure~\ref{fig:sectheta-plot} 
shows the averaged LDF for $\sec \theta \in [1.2,1.4]$ when the NKG assumption
is used. 
For comparison a hybrid derived average
LDF is shown too (see below for details). An energy dependent threshold effect 
is apparent at large radii and reflects upward fluctuations of signals close to the
trigger threshold of single tanks. 
In case of this NKG-like function with a free slope parameter, $\beta$, the
fit results for $E>\unit[10^{18}]{eV}$ are given in
Table~\ref{tab-beta}. 
The energy dependence of $\beta$ is shown in
Figure~\ref{fig:beta-vs-secTheta} and described by  
$\beta(E) = a(E) + b (\sec\theta -1)$, with        
$a(E) = 2.26 + 0.195 \log_{10}(E/{\rm EeV})$  and b = -0.98.
%
To quantify the quality of the fits the residuals, 
$(S - S_{\rm th})/\sigma_{\rm th}$, and their distributions
are computed. For a good LDF the residuals should scatter 
symmetrically around 0 with $\sigma = 1$. Means and standard
deviations of the residual distribution 
are used to compare different LDFs and are given in Table~\ref{tab-res}. For 
simplicity only residuals up to \unit[1500]{m} are taken into account, 
resulting in a variance smaller than the expectation value of 1 to avoid systematic
biases of upward fluctuating signals. 
The NKG-like function fits the data best, which can be seen from the 
smallest mean residuals and the
smallest residual variances. 

Complementarily to the SD analysis, a hybrid LDF analysis was performed. The
hybrid reconstruction exploits the independent knowledge of the core position
to determine the shower axis geometry and distance from each detector to the
shower core. A maximum likelihood fit (section \ref{sec-fit}) is used to
determine only S(1000) and the LDF slope parameter ($\beta$). A NKG-like
function was studied in the hybrid analysis. 

Hybrid triggers usually include a relatively large number of accidental
stations, which in the case of low multiplicity events could even outnumber
the number of stations that are part of the event making the identification of
accidentals and candidate stations a difficult task. Therefore, strict quality
cuts were imposed and only events with at least 6 triggered stations were
included in the analysis. That reduces the sample size, but at the same time,
selecting events with a large number of active stations, prevents biasing the
LDF slope due to signal fluctuations. Moreover, as the quality cuts
imposed 
\begin{wraptable}[6]{r}[0pt]{0.34\textwidth} 
  \vspace*{-3mm}
  \centering
  \scriptsize
  \begin{tabular} {|l|c|c|}
    \hline
    &   SD  &   Hybrid \\
    \hline
    Intercept $a$ &  2.24 $\pm$ 0.01  &  2.26 $\pm$ 0.17    \\
    Slope     $b$ & -0.98 $\pm$ 0.02  & -1.1  $\pm$ 0.3     \\
    \hline
  \end{tabular}
\vspace*{-4mm}
  \caption{Comparison between SD and Hybrid analysis on the parameterisation
    of the LDF slope (NKG-like function; $E>\unit[10^{18}]{eV}$). 
  }
\label{tab-beta}
\end{wraptable}
on the hybrid analysis are similar to 
those used on the SD-only
analysis 
the comparison is straightforward. 
Figure \ref{fig:beta-vs-secTheta} shows also how $\beta$ varies  with $\theta$
for hybrid events. 
Despite the
limited statistics of the hybrid sample the agreement between SD and Hybrid is
encouraging. 
The Hybrid data sub-sample is not large enough as to accurately
quantify the depencence of $\beta$ with energy at the time of writing.  
\vspace*{-2mm}
\section{Uncertainty in S(1000)}
\label{sec-s1000}
\begin{wrapfigure}[18]{r}[0pt]{0.48\textwidth}
\centering
\includegraphics[width=0.48\textwidth]{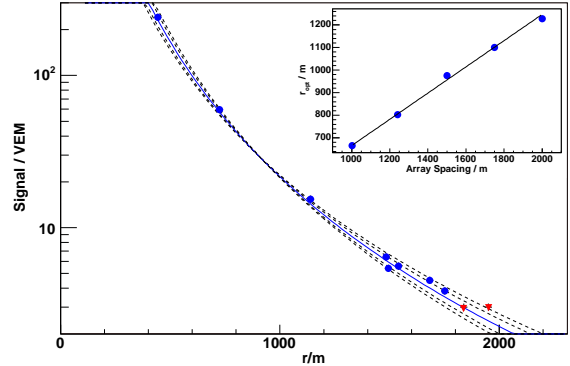}
\caption{ LDFs of an Auger SD event, analysed with
  different slope parameters. 
  Red points indicate stations with zero signal. 
The 
inset shows how the optimum ground parameter
  $r_{opt}$ varies with the array spacing.}\label{ropt} 
\end{wrapfigure}

High statistics are required to accurately describe the LDF, both to reduce
the statistical  and systematic uncertainties. Hybrid measurements
(though with much reduced statistics) are a useful tool to help identify sources of
systematic uncertainty, and it is certain that, as the Auger exposure increases,
the functional form of the LDF will evolve and become increasingly
accurate. Increasing accuracy will lead to much smaller uncertainties in the
reconstructed core position, 
but the  ground parameter, $S(1000)$ which is used to determine
the energy of the primary CR, is very robust to innacuracies in the LDF. It
can be shown that by measuring $S(r)$ at 1000 m,  fluctuations in the ground
parameter due to a lack of knowledge of the LDF are minimised. 
The result of analysing 
one SD event many times, whilst allowing the slope
parameter to vary by $\pm 8\%$ is shown in figure \ref{ropt}. This was chosen
as a reasonable value for the magnitude of the  shower-to-shower fluctuations, 
based on measurements made at Haverah Park where the precision was sufficient
to measure intrinsic shower-to-shower fluctuations \cite{coy1}. Analysing the
shower with  different assumed values for the slope parameter, results in a
shift in the reconstructed core position, but by choosing to measure the
ground parameter $S(r)$ at the point where the LDFs intersect, (at $\sim 1000$
m), 
the effect of the changing slope parameter is minimised. At this point
the ground parameter is independent of the LDF.
%
Analysing many showers in this way shows that $r_{opt}$, the optimum ground
parameter has very little dependence on  zenith angle, energy, or the form of
the LDF used to reconstruct the showers. For example, an analysis of $\sim
500$ Auger SD events with energies $10^{18.5}$ eV $< E < 10^{19}$ eV and zenith
angles $0^{\circ} < \theta < 60^{\circ}$ gives a distribution of $r_{opt} $
with mean 940 m and  a rms of 110 m. An analysis of simulated events at
$10^{20}$ eV gives a mean of 930 m and a rms deviation of 40 m. The prescence
of a saturated station (predominantly in vertical, high energy events)  tends
to push $r_{opt}$ out by several hundred metres, and after  an analysis of
many showers, at different zenith angle and energy,  S(1000) was chosen as a
robust ground parameter to measure all showers at. At 1000 m from the core the
mean uncertainty in $S(r)$ (across all events) is minimised, and furthermore,
for the few showers where $r_{opt}$ lies far from 1000 m,  the uncertainty in
S(1000) can easily be estimated. 

%
\vspace*{-2mm}
\section{Summary and Outlook}
\label{sec-conc}
The lateral distribution function of EAS observed using the 
Auger Observatory has been derived. Different functions have been tested and
it is concluded that an NKG-like LDF describes the data well. The dependence
of the function on atmospheric depth has been described. 
%
%
It should be emphasized that the global shower observables, like the
lateral distribution of particles, are  not affected by the geomagnetic field
for zenith angles $\theta < 70^{\circ}$. However, for the case of very
inclined showers which are dominated by muons, the density at ground is
rendered quite asymmetric by the geomagnetic field and the LDF approach is
not longer valid~\cite{ref:ave-astroph,ref:hillas}.


%

\end{document}